\documentstyle[12pt,aaspp4]{article}

\lefthead{Secker et al.}
\righthead{Radiopanspermia Revisited}

\begin{document}

\title{Astrophysical and Biological Constraints on Radiopanspermia}

\author{Jeff Secker}
\affil{ Washington State University, Program in Astronomy, Pullman, WA 99164-3113 USA}
\authoremail{secker@delta.math.wsu.edu}

\author{Paul S. Wesson and James R. Lepock}
\affil{University of Waterloo, Department of Physics, Waterloo, ON  N2L 3G1 Canada} 

\begin{abstract}
We have carried out a series of calculations involving bacteria and
viruses embedded in dust grains, which are ejected from our solar
system by radiation pressure, and travel through space to other star
systems.  Under many conditions, this kind of panspermia is
impractical, primarily because the ultraviolet (UV) radiation of the
present Sun inactivates the micro-organisms.  However, if the
organisms are shielded by an absorbing material like carbon, and if
ejection takes place in the late-Sun (red-giant) phase of a
one-solar-mass star like our Sun, there is a significant probability
that these micro-organisms can reach another star system alive (i.e.,
with only sub-lethal damage from UV and ionizing radiation).  In
addition to panspermia with viable micro-organisms, we note that it is
possible to seed the Galaxy with inactivated ones, whose DNA and RNA
fragments may provide the initial information necessary to start
biological evolution in favorable environments.
\end{abstract}

\section{Introduction} 

The panspermia hypothesis proposes that life originated on a planet
other than Earth, and that this life (in the form of primitive
biological micro-organisms) was introduced onto the Earth after it
traveled through interplanetary or interstellar space.  Thus if
panspermia is a viable mechanism for disseminating life throughout the
galaxy, the origin of life on Earth is intrically connected to the
origin and abundance of life on other planets in the Galaxy.
Recently, the panspermia hypothesis and related areas of research have
received considerable attention by Hoyle, Wickramasinghe and their
collaborators (e.g., Hoyle \& Wickramasinghe 1979, 1981, 1982, 1983;
Jabir, Hoyle \& Wickramasinghe 1983; Hoover et al. 1986; Hoyle,
Wickramasinghe \& Al-Mufti 1986).  For a detailed review of
panspermia, we refer interested readers to the literature (e.g.,
Shklovskii \& Sagan 1966, p. 206; Crick \& Orgel 1973; Tobias \& Todd
1974, p.238; Horneck \& Brack 1992; Horneck 1995).

The panspermia hypothesis can be separated into distinct classes,
based primarily upon the method used to transport the micro-organisms
through the space environment: among these are {\em lithopanspermia,
radiopanspermia and directed panspermia}.  Lithopanspermia refers
to micro-organisms which are carried to the Earth's surface within
meteorites. From probablistic arguments, lithopanspermia is limited to
interplanetary travel within an individual solar system (e.g., Tobias
\& Todd 1974, and references therein).  Directed
panspermia refers to the hypothesis that life was seeded on the Earth,
deliberately, by an intelligent civilization on another planet (Crick
\& Orgel 1973).  In this case, the transport mechanism would be a
spacecraft constructed specifically for this task.  On the other hand,
radiopanspermia refers to single exposed micro-organisms, which are
small enough to be accelerated to high velocities by solar radiation
pressure, then exit the solar system and traverse interstellar space
(Arrhenius 1908).  Provided that some enormous number of
micro-organisms were ejected from a solar system, it is conceivable
that a small fraction may intersect a hospitable planet in a new star
system, after $\sim 10^6$ years in interstellar space.  In order for
this radiopanspermia to effectively seed life on the new planet, some
small fraction of the micro-organism must survive the radiation damage
and the harsh environment of free space.  However, it is known that
exposed micro-organisms are immediately (i.e., within a few days)
inactivated by solar ultraviolet (UV) radiation, even before they
leave the original solar system.

We (Secker, Lepock \& Wesson 1994; hereafter SLW94) have carried out a
series of calculations on a plausible variant of the radiopanspermia
hypothesis, which combines the radiation shielding inherent to lithopanspermia
with the transport mechanism of radiation pressure.  Herein micro-organisms
embedded in cosmic dust grains are ejected from their original solar
system by radiation pressure, traverse space, and arrive in another
solar system, where after deceleration and deposition to a new planet
is possible.  However, we restrict the grain properties to the region
of the radius-density parameter space for which the repulsive
force of the Sun's radiation pressure dominates over the attractive
force of the Sun's gravity, as required for distribution by radiation
pressure.

Our motivation for this analysis was to see how far known
astrophysical conditions can help spread biological micro-organisms
through the Galaxy.  Our calculations concentrated on the ejection of
the bacteria and viruses from a solar system like ours by radiation
pressure and the survival of these against harmful UV and ionizing
radiation.  The effects of vacuum and low temperatures are not as bad
as might be expected, and it is the UV radiation of the host stars
that presents the greatest biological hazard (e.g., Glenister \& Lyon
1986; Horneck 1982; SLW94; Brack \& Horneck 1995).  For other aspects
of the panspermia hypothesis, such as the departure of microbes from a
planet into space (e.g., Melosh 1988; Moreno 1988; Hoyle,
Wickramasinghe \& Al-Mufti 1986), and a discussion of other
potentially-damaging processes (e.g., SLW94; Horneck \& Brack 1992),
we refer the reader to the literature.
 
Our main conclusion will be that panspermia with living
micro-organisms may be viable, but only under a certain set of
circumstances wherein they are shielded from radiation by the
absorbing mantle of a carbonaceous dust grain and ejected into space
in the red-giant (late) stage of a one-solar-mass star's life.
However, we were also interested to see if we could alleviate the
problem that the origin of life by purely random chemistry, without
the intervention of biological information from elsewhere, may be very
unlikely (Wesson 1990).  We will note as an addendum that panspermia
with inactivated micro-organisms, even though it may sound
unpalatable, may be viable from the informational point of view.
 
\section{Conditions for Panspermia}

We start by considering the ejection from our solar system of a particle
which we later identify with a dust grain containing either a bacterium or a
virus.  Then we will consider the radiation dosage to which the organism is
subjected in order to decide on its survival.
 
To first order, the net force on a particle is the difference between that
due to radiation pressure and that due to gravity.  Let $L_{\odot}$ and
$M_{\odot}$ equal the Sun's luminosity and mass, $r$ equal the distance
between the grain and the Sun, $G$ equal the gravitational constant, $c$
equal the velocity of light, and $R$ and $m$ equal the radius and mass of the
particle. Then the force due to gravity and the classical expression
for radiation pressure are
 
\begin{equation}
F_{g} =  \frac{GM_{\odot}m}{r^{2}}
\end{equation}
 
\begin{equation}
F_{p} =  \frac{L_{\odot} \pi R^{2}}{(4 \pi r^{2})c}.
\end{equation}
 
\noindent Here we have assumed a radiation efficiency factor which is of
order unity, that $F_p$ and $F_g$ are the dominant forces, and we
assume that this classical expression adequately describes the
ejection of carbonaceous dust particles (Burns, Lamy \& Soter 1979;
SLW94).  The acceleration of the particle is then given by
 
\begin{equation}
\frac{d^{2}r}{dt^{2}} =  \frac{C}{r^{2}}, \ \ {\rm where} \ \
C \equiv \left( \frac{L_{\odot}R^{2}}{4mc} \right)
-GM_{\odot}.
\end{equation}
 
\noindent The velocity at distance $r$, assuming $v(r=r_i)=0$ at $t=t_0$, is
 
\begin{equation}
v(r) = \left( \frac{2 C}{r_{i}} - \frac{2C}{r} \right)^{1/2}.
\end{equation}
 
\noindent The time $t$ corresponding to distance $r$ is obtained by
integrating (4) (with $r=r_i$ at $t=t_0$) to give
 
\begin{equation}
t(r) = \left( \frac{r_{i}}{2C} \right)^{1/2} \left\{ r(1 - r_{i}/r)^{
1/2}+ r_{i} \ln \left[(\frac{r}{r_{i}})^{1/2} +
(\frac{r}{r_{i}}-1)^{1/2} \right] \right\}+t_{0}.
\end{equation}
 
For ejection, we require that $C > 0$ in (3), so that only small,
low-density particles can be ejected.  When a particle leaves the
solar system, it will do so with a terminal velocity $v_{t} =
(\frac{2C}{r_{i}})^{1/2}$ given by (4).  And after a certain time to
exit the solar system (given by 5), it will travel for a further
period determined by $v_t$ and the distance to a neighboring star
system.  The particle's travel between neighboring stars will either
be a random walk (a mean-free path with $l \simeq 10^9$ cm, due to
collisions with neutral hydrogen), or it will be bound-up in and
travel with an interstellar cloud (Weber \& Greenberg 1985;
SLW94). The velocity of molecular clouds is about 10 km s$^{-1}$
(Spitzer 1978; p.44), roughly equivalent to the terminal velocity of
the particles as they exit a solar system. We therefore calculate the
particle's interstellar travel time using their exit velocity.  Within
about 20 light years there are more than 50 stars, and it is this
distance we adopt for interstellar travel.  Finally, {\em it is the
relative velocity of the particle with respect to that of the host
planet that governs the energy of impact with the atmosphere}, but the
deceleration caused by the second star's radiation pressure is also a
factor. That is the panspermia transport mechanism as we envision it.
 
We (SLW94) have carried out a detailed set of calculations using
different parameters for the preceding relations, which we only
summarize here.  For particles with sizes on the order of $10^{-6}$ cm
and densities in the range $0.9-2.2$ gm cm$^{-3}$, and considering
both the current Sun and a one-solar-mass red-giant star, times to
exit the solar system are on the order of 35 years and times to reach
a neighboring star system are of the order $10^6$ years; Tables 1 and
2 of SLW94 describe the full results.  Since these stars have a
significant component of radiation that would harm a micro-organism, we
need to inquire about survival.  In SLW94, the radiation calculations
were considered in three parts: the solar UV radiation, the solar
ionizing radiation, and the interstellar radiation field.

The Sun's UV radiation field can be conveniently modeled as part of a
Planck function with an effective temperature of $T_{\odot}
\simeq 5785$ K.  Thus, if $h$ is Planck's constant and $k$ is
Boltzmann's constant, the flux at a distance $r$ from the Sun
(radius $R_{\odot}$) at wavelength $\lambda$ is given by
 
\begin{equation}
F(r,\lambda) = \frac{R_{\odot}^{2}}{r^{2}}\left(\frac{2 \pi hc^{2}}
{\lambda^{5}(e^{\frac{hc}{\lambda kT_{\odot}}}-1)} \right).
\end{equation}
 
\noindent Note  that this black-body approximation overestimates
the UV radiation emitted by the Sun.  As described by Tobias \& Todd
(1974; p.241), about half of the UV radiation emitted by the Sun in
the range $2500-2900$ \AA \ is absorbed by its corona.  The ionizing
component of the Sun's radiation, $0 < \lambda < 1600$ \AA, was
modeled using polynominal functions fit to experimentally-measured
data (see SLW94).  Then the total dose accumulated within the radius
range $r_i$ to $r$ (i.e., the Solar System), with the wavelength
integral taken over $0-3000$ \AA, is given by
 
\begin{equation}
D(r) = \left( \frac{2(1-r_i/r)}{r_i C} \right)^{1/2} 
\int  F(\lambda) d\lambda.
\end{equation}
 
In interstellar space, the intensity of radiation is many orders of
magnitude less than it is in the vicinity of the Sun.  The
interstellar radiation field is provided by an ensemble of stars at
large distances, and this is best modeled as a Planck function with a
temperature of $10^4$ K, multiplied by a geometric dilution factor of
$10^{-14}$ (SLW94).  In this case, the total radiation dose received in
interstellar space $D_{\rm is}$ is calculated to be the interstellar
travel time $t_{\rm is}$ multiplied by the integral over the
wavelength range $0-3000$ \AA \ of the model Planck function.  To
determine the survival of an organism, we worked out total doses
taking all components of radiation into account, and considered the
radiation resistance of bacteria and viruses.

Three different micro-organisms were considered in these
calculations.  The {\em Micrococcus radiophilus} is the most
radiation-resistant bacteria known at this time, and it is therefore a
logical candidate for this radiopanspermia.  The {\em Staphylococcus
minimus} is a very common bacteria which is much smaller than the {\em
Micrococcus radiophilus}. As well, the virus we considered combined
properties of both the {\em T1 Bacteriophage} and the {\em phage
C-36}.  Refer to SLW94 for a discussion of the characteristics of
these micro-organisms.  The survivability of a given micro-organism was
quantified using the relevant inactivation cross section $\sigma$, in
terms of which the number $N$ surviving from an initial number $N_0$
after dose $D$ is
 
\begin{equation}
N = N_0 \exp^{-\sigma D}.
\end{equation}
 
\noindent Here $\sigma$ depends in general on $\lambda$, and we used 
the appropriate polynominal functions fit to the
experimentally-measured inactivation cross sections to interpolate
between wavelength.  (Note however, that these behave somewhat
differently in the low-pressure, low-temperature environment of space;
e.g., Levine \& Cox 1963; Ashwood-Smith, Copeland \& Wilcockson 1968;
Tobias \& Todd 1974; Weber \& Greenberg 1985).  Using parameters for
the current Sun, computing the total dose as described above (assuming
the exposure in the new system is the same as in the old), and
adopting these inactivation cross sections, leads to values of $N/N_0$
consistent with zero.  {\em We find that the Sun's UV radiation is
considerably more harmful than its ionizing radiation, and it is so
intense at the present time that it effectively inactivates all
exposed micro-organisms}.

This situation might be avoided if the micro-organisms are embedded in
dust grains.  This might be a natural thing, depending on how they are
put into space, through UV processing of a thin surface skin of
organic matter, or through interactions and accretion of carbon-rich
interplanetary dust particles (Chyba \& Sagan 1988).  And we know
there are grains of the appropriate size in space (Mathis, Rumpl \&
Nordsieck 1977).  We therefore considered the survival of bacteria and
viruses shielded by thin films (spherical shells) of various materials
known to exist in space.  Of these thin films, water ice is common but
does not help because it transmits too much UV radiation. On the
contrary, carbonaceous material is an effective blocker.  However, we
now have another factor to contend with: a mantle that protects the
organism also increases the mass of the particle to be ejected from
the solar system, and it is difficult to get the organism-plus-grain
out into interstellar space with the radiation pressure of the Sun as
it is now.  Thus, even shielded micro-organisms do not make panspermia
work with the present Sun.
 
We should not conclude that panspermia is impossible, for in the late,
red-giant stage of a star like the Sun its luminosity increases.
Also, it is natural to consider the ejection of micro-organisms in the
later stages of a stars life, because by then evolution will have had
longer to run on any planets which may be around the star, increasing
the chances of there being biological material to distribute. For
these reasons, we examined the ejection of micro-organisms during the
red-giant phase of a one-solar-mass star; the results are given in
Tables 3 and 4 of SLW94.  Our essential conclusion is that panspermia
is viable in this case: given shielding by carbonaceous material with
thicknesses on the order of 0.25-0.68 $\mu$m, the total doses are so
low that the survival fraction $N/N_0$ is close to unity (i.e. $0.95 -
1.00$) for all three micro-organisms considered here.  {\em That is,
panspermia is viable, provided one considers micro-organisms shielded
within carbonaceous dust grains and ejected from stars like the Sun in
their late stage of life}.  An interesting aspect of these
calculations concerns the total UV doses received by the two bacteria,
after attenuation by the thin carbonaceous layer. These are $D \simeq
6\times 10^2$ erg/mm$^2$ for the {\em Micrococcus radiophilus}, as
compared to $D \simeq 6\times10^{-11}$ erg/mm$^2$ for the {\em
Staphylococcus minimus}.  Thus the smaller micro-organisms can accrete
a thicker coating of attenuating carbonaceous material, and still be
within the mass-radius range for acceleration by radiation pressure.

A final comment with respect to the radiation exposure calculations is
worthwhile.  There is a fundamental difference between the
radiation-induced {\em damage} to bacterial DNA and RNA and its
subsequent {\em repair}.  The total radiation dose imparted to a given
micro-organism occurs continuously over a period of about $10^6$
years.  However, micro-organisms cannot repair sub-lethal DNA and RNA
damage in the extremely cold and desiccated conditions of interstellar
space.  Thus all structural damage accumulates until the
micro-organisms arrive in a hospitable environment: provided that the
total accumulated damage is below the lethal level, the
micro-organisms can at this point repair the damage.  This
damage-accumulation effect has been accounted for in our calculations.
A similar effect concerns viral DNA and RNA, which also requires a
hospitable environment for repair, in this case a specific, living
host which to infect.

\section{Discussion}

We have taken a new look at the old idea of panspermia, examining the
astrophysical conditions under which an organism can leave one solar
system under the influence of radiation pressure and gravity, as well
as the biological conditions under which it can survive radiation
damage and arrive at another solar system in a viable state.  While
work only scratches the surface of this subject, we conclude that the
traditional idea of radiopanspermia is valid if micro-organisms
(bacteria and viruses) are shielded inside grains whose material
blocks significant UV radiation, and are ejected into space in the
late stages of a (one-solar-mass) star's life.  Coupled with recent
discoveries supporting other aspects of panspermia (Horneck \& Brack
1992; Horneck 1995), our results suggest that the probability for life
in any given solar system has increased.  This finding is relevant to
the current search for extraterrestrial intelligence ({\sc SETI};
e.g., Horowitz \& Sagan 1993; Tarter 1994), which is indeed a
worthwhile endeavor.

As did others, (e.g., Horneck 1982; Greenberg \& Weber 1985), we
determine that the damage caused by solar UV radiation (i.e., during
the time spent in either solar system) is the most deleterious to the
micro-organisms, and which limits this panspermia.  Our calculations
reveal that the dominant factor in survival against UV radiation is
not the radiation resistance of the micro-organism, but it is
thickness of the carbon layer which attenuates the UV radiation.  As
illustrated in the above calculations, the maximum possible thickness
of this layer depends upon the diameter of the micro-organism, and on
the luminosity-to-mass ratio of the star.

Considering the origin and spread of life, it is worthwhile to extend
the scope of this radiopanspermia from living (or potentially viable)
micro-organisms to any form of biological material including dead
(inactivated) micro-organisms.  The rationale for this has simply to
do with information. If the original atmosphere of the Earth was
composed largely of carbon dioxide as opposed to the reducing
atmosphere commonly hypothesized, then the organic compounds necessary
for the origin of life would be more difficult to generate.  A
panspermia that distributes biological material in the form of
fragmented DNA, RNA and protein bound up in inactivated bacteria and
viruses supplies information.  This biological material is available
for replication, and if it is introduced to a hospitable environment,
it may enhance the chance that life will evolve there, and could
possibly explain the (apparently) rapid evolution of early life on
Earth.  With this approach, the seeding of life through the Galaxy via
radiopanspermia need not rely solely on living organisms, as the
dispersal of inactivated biological material such as fragmented DNA
and RNA is relatively easy to accomplish.

\acknowledgments

This research was supported by grants to P.S.W. and J.R.L.  from the
Natural Sciences and Engineering Research Council of Canada.  We would 
like to thank an anonymous referee for helpful comments.

\end{document}